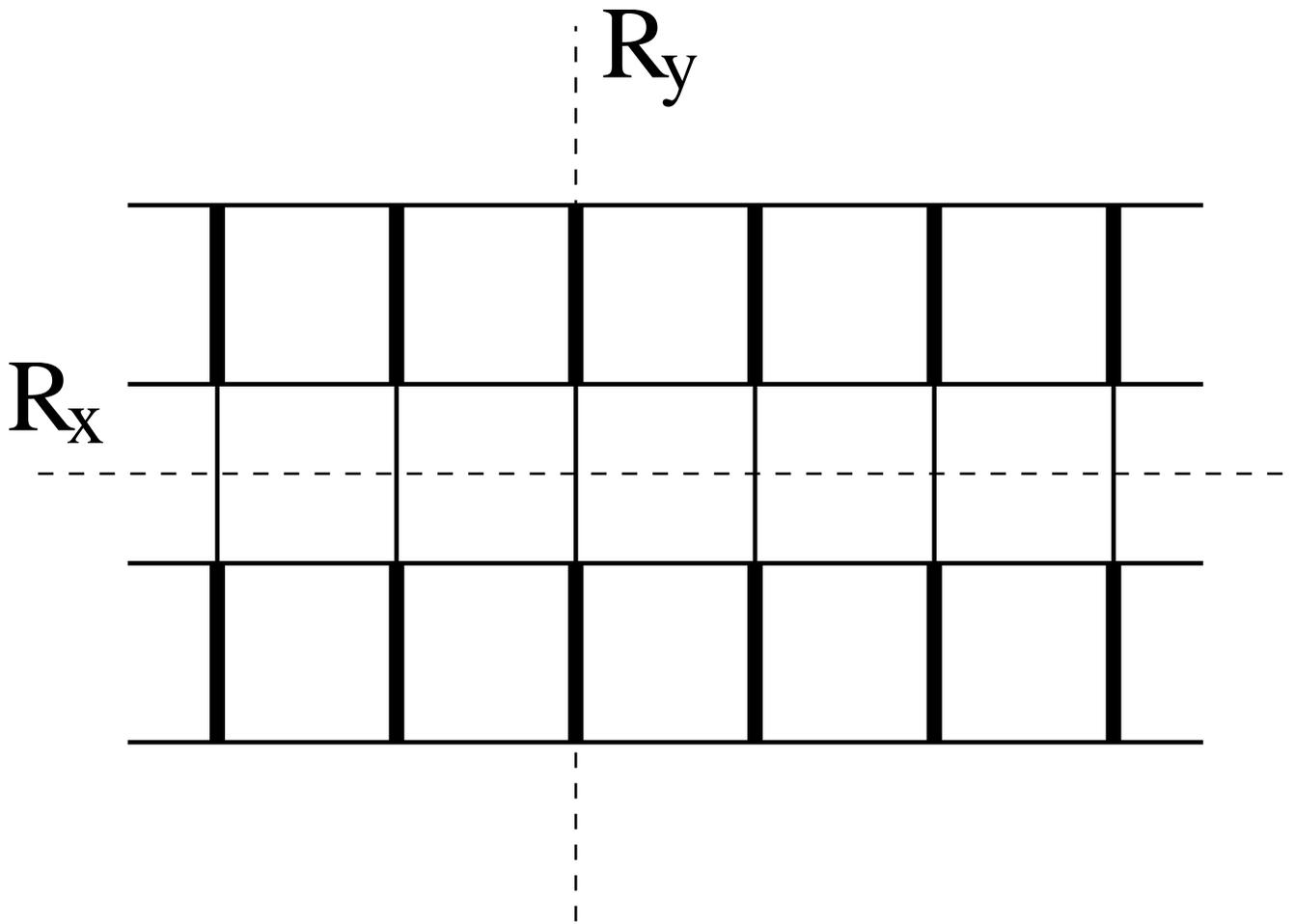

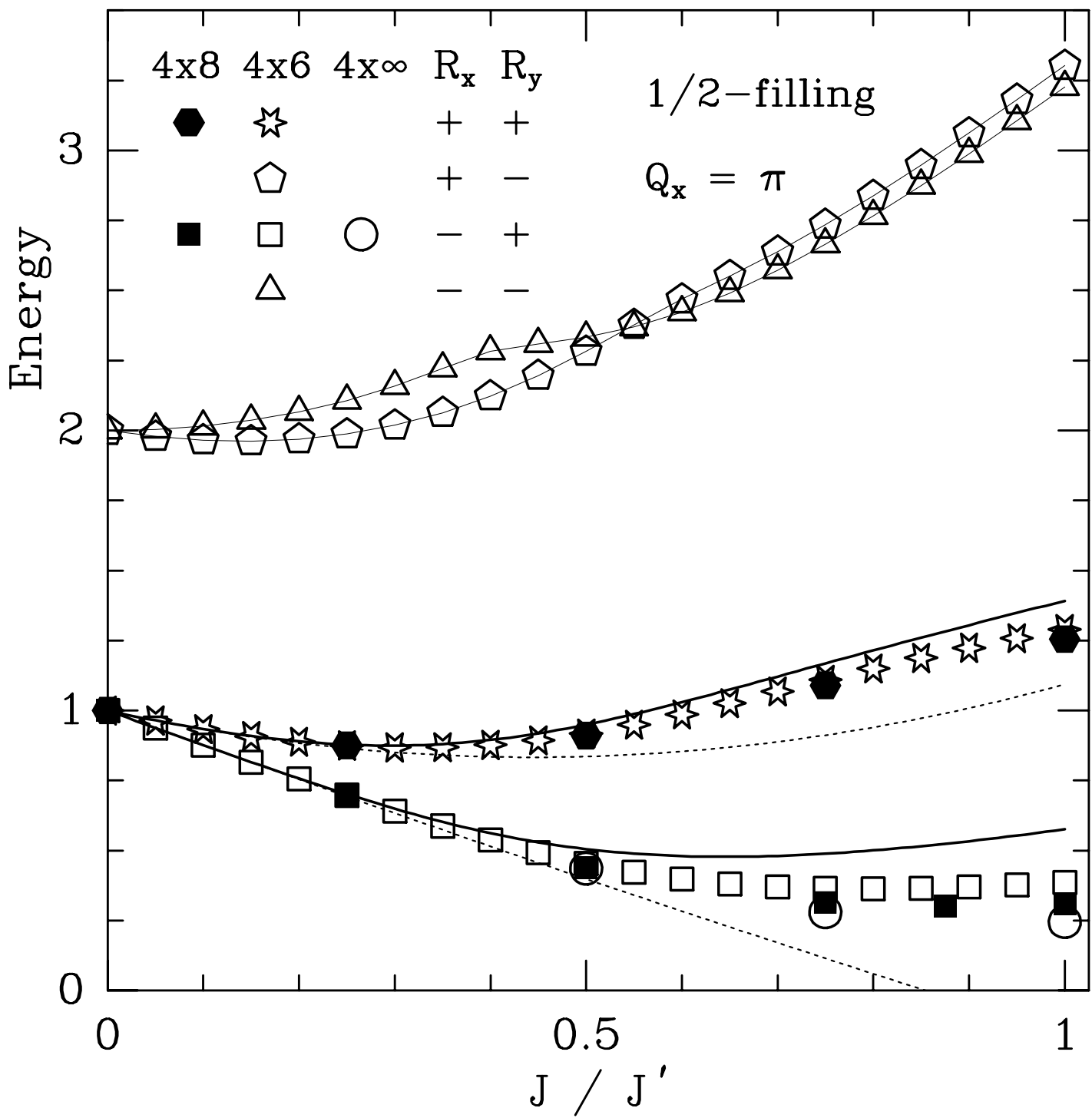

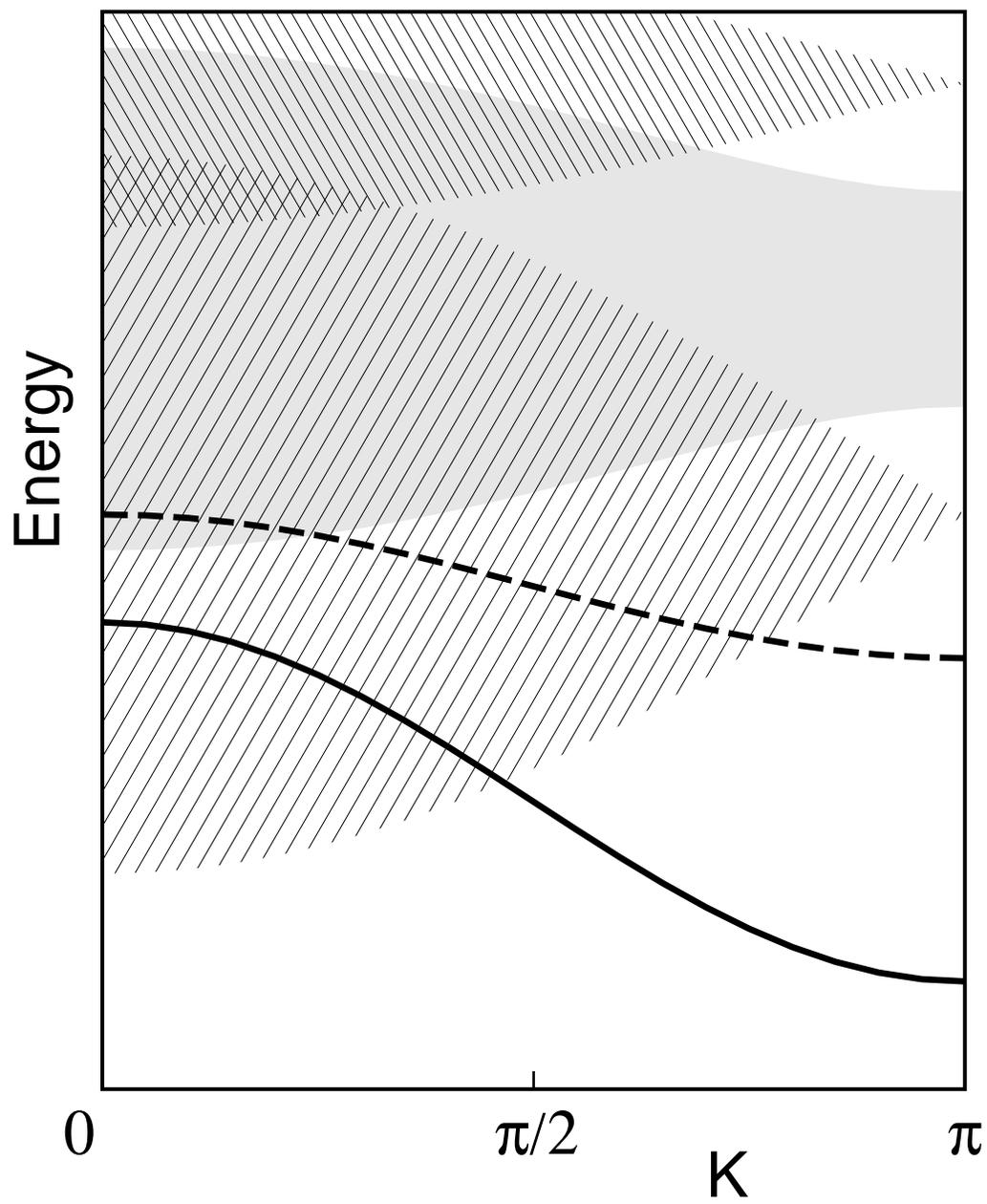

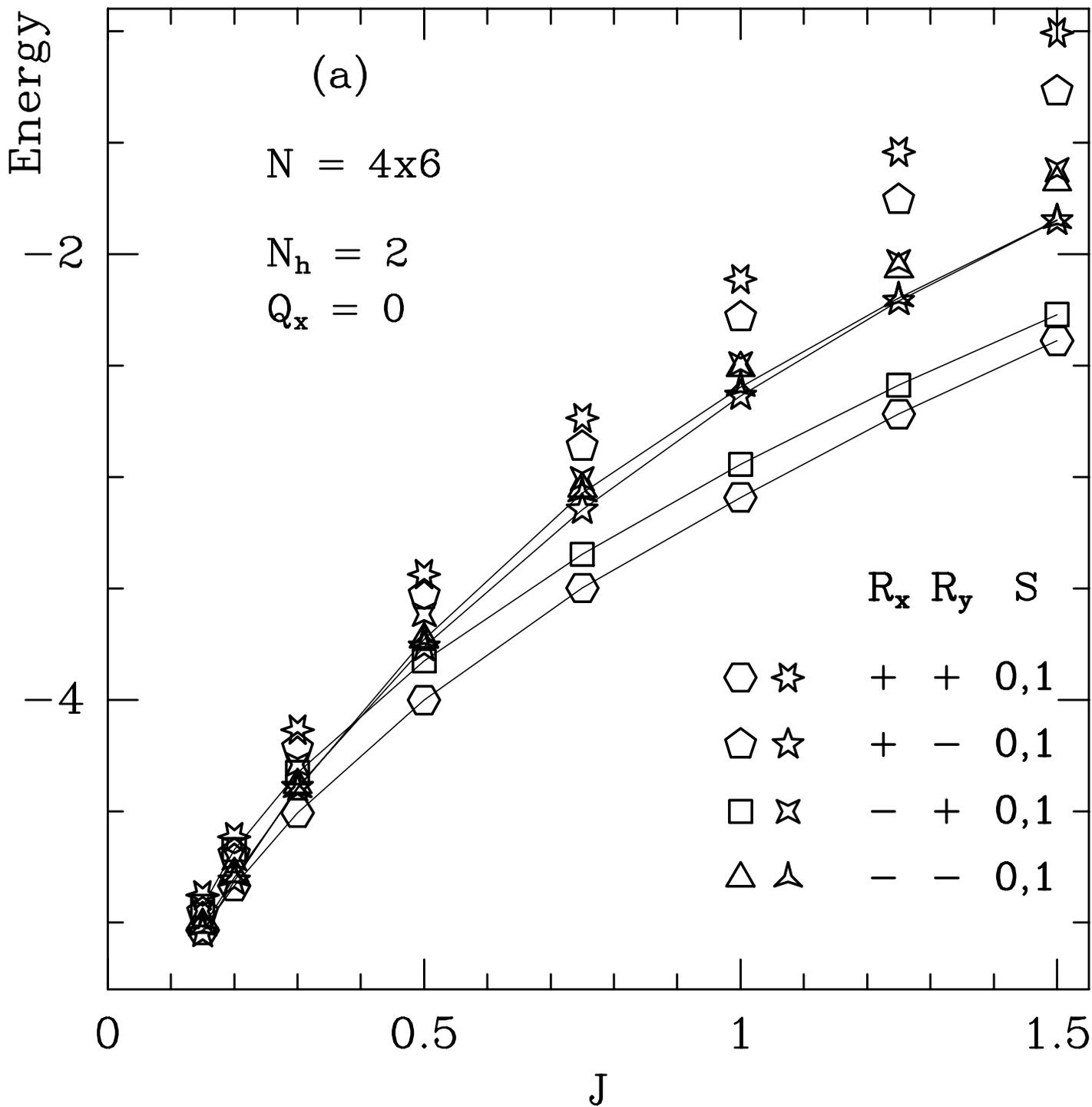

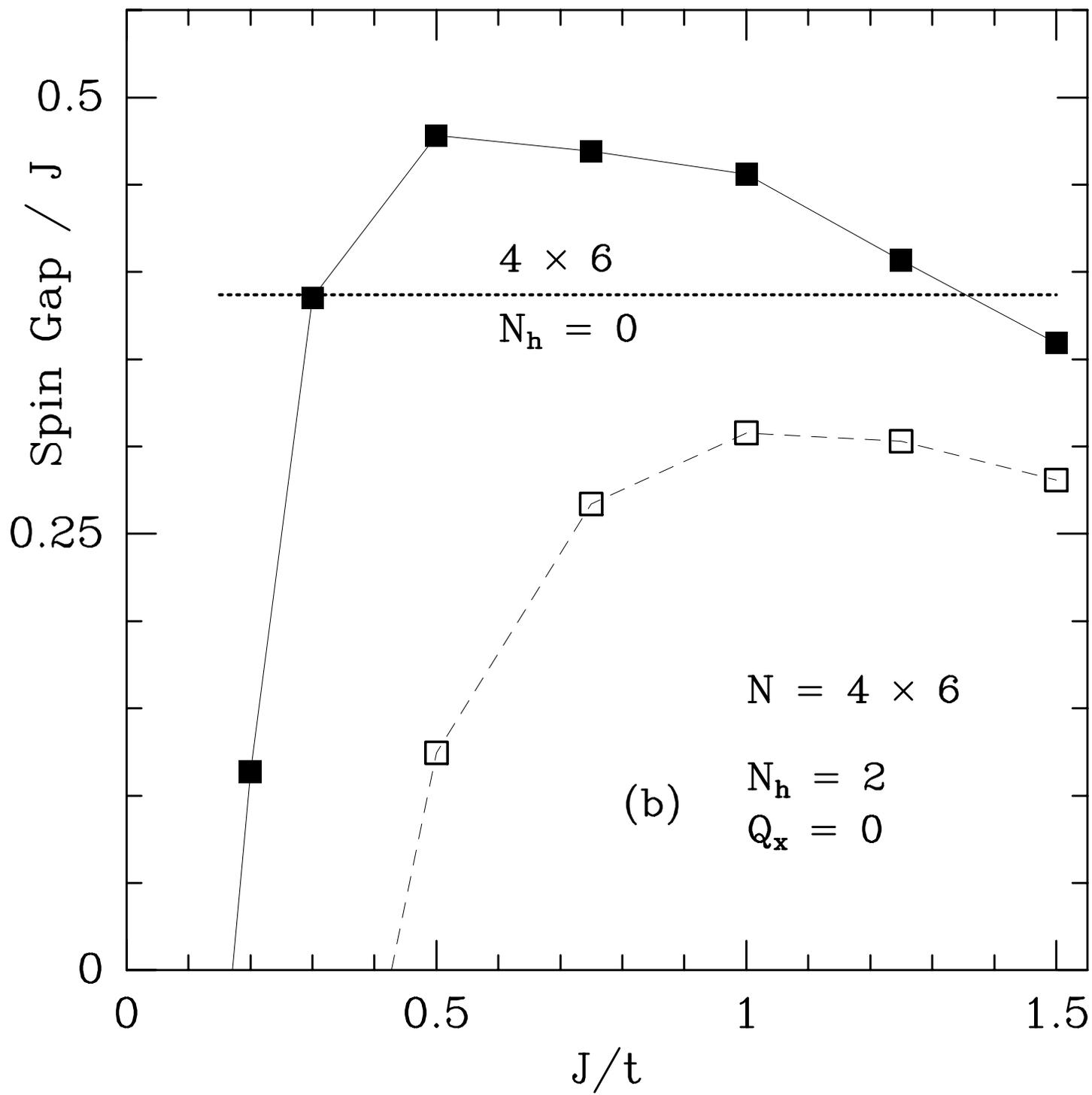

# Spin Gaps in coupled $t$–$J$ ladders

D. Poilblanc[1], H. Tsunetsugu[2,3] and T.M. Rice[2]

[1] *Lab. de Physique Quantique, Université Paul Sabatier, 31062 Toulouse, France*

[2] *Theoretische Physik, ETH-Hönggerberg, 8093 Zürich, Switzerland*

[3] *Interdisziplinäres Projektzentrum für Supercomputing, ETH-Zentrum, 8092 Zürich, Switzerland*

(Received)

## Abstract

Spin gaps in coupled $t$-$J$ ladders are investigated by exact diagonalization of small clusters up to 4×8 sites. At half-filling, the numerical results for the triplet excitation spectrum are in very good agreement with a second order perturbation expansion in term of small inter-ladder and intra-ladder exchange couplings between rungs ($J/J'$<0.25). The band of local triplet excitations moving coherently along the ladder (with momenta close to $\pi$) is split by the inter-ladder coupling. For intermediate couplings finite size scaling is used to estimate the spin gap. In the isotropic infinite 4-chain system (two coupled ladders) we find a spin gap of $0.245J$, roughly half of the single ladder spin gap. When the system is hole doped, bonding and anti-bonding bound pairs of holes can propagate coherently along the chains and the spin gap remains finite.

Strongly correlated systems described by $t$-$J$ or Hubbard models in confined geometries have recently been the subject of many investigations. One motivation is the interest in the role of the dimensionality. Varying the system size in a transverse direction often enables one to study dimensional cross-overs. Another motivation is the relevance to real physical systems where the electrons are confined in some direction. It was found[1] that some cuprates ($Sr_{2n-2}Cu_{2n}O_{4n-2}$) consist of parallel multi-ladders with $n$-chains weakly coupled to each other. It has been argued[2] that such a system would be a prototype of a resonating valence bond (RVB) spin liquid when $n$ is even. Another class of multi-ladder compounds are $La_{4+4n}Cu_{8+2n}O_{14+8n}$.[3] In particular, both $Sr_6Cu_8O_{14}$ and $La_2Cu_2O_5$ consists of coupled 4-chain ladders.

Most theoretical work so far has been on the single ladder (double chain). At half-filling (Heisenberg single ladder), the ground state (GS) can be viewed as a liquid of singlet rungs. The lowest triplet excitation is obtained by exciting a singlet rung to a triplet which costs an energy gap proportional to the exchange coupling. Strong coupling perturbation expansions[4,5], mean-field (MF) approaches[6] and numerical methods[7,8] determine the gap in the spin excitation spectrum to be $\approx 0.5J$ in the isotropic coupling limit. When holes are doped into the system they form bound pairs on the rungs (to minimize the magnetic cost) which can then propagate along the ladder.[9] In most physical cases the coupling between the ladders (double chains) becomes relevant. It is then crucial to investigate the role of the ladder coupling. MF theory suggests[6] that the spin gap for coupled ladders is strongly reduced, however, its validity needs to be tested by numerical calculations. The aim of this paper is to study numerically double ladder clusters (4-chain systems). In the parameter space where the perturbation expansion breaks down, finite size scaling analysis can however give valuable results. The spin gap and the correlation length of the infinite double ladder can then be estimated. Our numerical data also suggest the existence of hole pairs and a spin gap away from half-filling.

The $N$-ladder $t$-$J$ Hamiltonian is,

$$\mathcal{H} = J' \sum_{\alpha=1}^{N} \sum_{j} (\mathbf{S}_{j,2\alpha-1} \cdot \mathbf{S}_{j,2\alpha} - \tfrac{1}{4} n_{j,2\alpha-1} n_{j,2\alpha})$$
$$+ J \sum_{\beta,j} (\mathbf{S}_{j,\beta} \cdot \mathbf{S}_{j+1,\beta} - \tfrac{1}{4} n_{j,\beta} n_{j+1,\beta})$$
$$+ J \sum_{\alpha=1}^{N-1} \sum_{j} (\mathbf{S}_{j,2\alpha} \cdot \mathbf{S}_{j,2\alpha+1} - \tfrac{1}{4} n_{j,2\alpha} n_{j,2\alpha+1})$$
$$- t \sum_{j,\beta,s} P_G (c^\dagger_{j,\beta;s} c_{j+1,\beta;s} + c^\dagger_{j,\beta;s} c_{j,\beta+1;s} + \mathrm{H.c.}) P_G, \quad (1)$$

where most notations are standard. The $\beta$ (=1,...,$2N$) indices label the $2N$ legs of the $N$ ladders (oriented along the $x$-axis) while $j$ is a rung index ($j$=1,...,$L$). Since the $\alpha^{th}$ ladder contains the two legs labeled by $2\alpha-1$ and $2\alpha$ the sum in the $J'$ term is restricted to the isolated rungs of the $N$ ladders. The two $J$ terms correspond to the intra- (along $x$) and inter-ladder (along $y$) magnetic couplings respectively. Finally, the $t$ term describes the hole propagation between all nearest neighbor links. Periodic boundary conditions are used in the $x$- (chain) direction. Below we will deal mainly with the double ladder ($N$=2) depicted in Fig. 1. Most conclusions are generic and could be generalized to a larger $N$.

Small $4 \times L$ ($L \leq 8$) clusters were diagonalized by a Lanczos method by using spin inversion, $S^z_{j,\alpha} \rightarrow -S^z_{j,\alpha}$, (for even particle number) to distinguish $S_{\mathrm{tot}}$=0, 1 as well as full spatial symmetries, and the GS energies were obtained in each symmetry sector. Finite size scaling arguments are then used to extrapolate to $L \rightarrow \infty$.

Let us first consider the half-filled case. Hereafter $J'$ sets the energy scale. When $J$=0, $\mathcal{H}$ reduces to isolated rungs and the spin excitations can be constructed by exciting singlet rungs to triplets each costing an excitation energy, $J'$. The excitation spectrum consists of equally spaced levels. Once the rungs have an intra-ladder coupling $J$, the triplets can propagate, form a coherent band, and the spin gap is reduced. As in the single ladder case,[6] the lowest energy is at a momentum $K_x$=$\pi$ (in the chain direction). The inter-ladder coupling we expect will split the single ladder triplet. This splitting can be calculated perturbatively in $J$. Up to the 2$^{\mathrm{nd}}$ order the spin gap of a single ladder is $J'-J+\tfrac{1}{2}J^2$.[4,5] The inter-ladder coupling in the same order gives a splitting between the bonding ($B$) and antibonding ($A$) combination so that the gaps at $K_x$=$\pi$ are

$$1 - \tfrac{5}{4}J + \tfrac{3}{32}J^2 \ (B), \quad 1 - \tfrac{3}{4}J + \tfrac{27}{32}J^2 \ (A). \tag{2}$$

The true spin gap, i.e. the smallest excitation energy to a triplet state, is the $B$-mode. The single band of the single ladder gives then rise to $B$- and $A$-excitation bands in the double ladder.

Triplet excitation energies (measured from the GS energy) at $K_x=\pi$ for various $4\times L$ clusters ($L=4$, 6, 8) are shown in Fig. 2. Excited states are classified according to their parity w.r.t. the reflections along the $x$-axis or $y$-axis (see Fig. 1). As shown in Fig. 2, the numerical data agree remarkably well with the 2$^{\text{nd}}$ order perturbation when $J\leq 0.25$. In this case the spin-spin correlation length $\xi_S$ is small (of a few lattice spacings), finite clusters with $L\gtrsim\xi_S$ can give accurate results. The $B$- and $A$-excited triplet states are odd and even under a reflection $R_x$. On the other hand, as shown in Fig. 2, the excited states involving one (energy $\sim 1$) or two (energy $\sim 2$) triplet rungs differ by opposite symmetries under a reflection $R_y$ w.r.t. some $y$-axis along one rung (see Fig. 1). When $J$ becomes larger 2$^{\text{nd}}$ order perturbation obviously breaks down and one has then to rely on numerical results. As for the Haldane $S=1$ chain with *periodic* boundary conditions we expect the spin gaps $\Delta(L)$ in the $4\times L$ clusters to scale as[10]

$$\Delta(L) \approx \Delta(\infty) + ae^{-L/\xi}, \tag{3}$$

apart from (presumably) small logarithmic corrections in the argument of the exponential. Least square fit of the data of Fig. 2 shows that the form (3) is well satisfied. This contrasts with the (larger) leading finite size corrections ($\sim 1/L^2$) expected for multi-chains with *open* boundary conditions.[11] The best parameters and the spin gaps for the two largest clusters are shown in Table I. Note that physically the spin-spin correlation length $\xi_S$ should in principle be directly proportional to $\xi$. Then, from Table I we expect e.g. $\xi_S$ to increase by only a factor of order 1.7 when $J$ increases from $J=0.5J'$ to the isotropic (in real space) case $J=J'$. This short correlation length, $\xi<L/2$, may in turn justify using the scaling (3). From these scaling arguments we obtain a value of $0.245\pm 0.02J$ for the spin gap at $J=J'$; a value much larger than that ($0.12J$) obtained in a MF theory.[6] This difference is due in part to dynamical polarizations. A typical process is that two neighboring singlet rungs virtually polarize into triplets, lowering the GS energy, but the energy gain of an existing triplet is smaller because of fewer polarization channels. Thus, the spin gap is enhanced by polarization processes ignored in MF theory. A smaller value of the spin gap, $0.209J$, was recently obtained by a density-matrix renormalization-group method.[8] The origin of the discrepancy is not clear yet.

Delocalization along the chain direction of the $B$ and $A$ local triplet excitations gives rise to two split triplet bands (say, single "magnons") with (roughly) cosine-like dispersions (in k-space) and minima located at $K_x = \pi$. Their bandwidths are directly proportional to the finite hopping amplitude of the triplet bond to hop on neighboring sites. From our previous numerical analysis including the estimated spin gaps and neglecting interactions among magnons, we can qualitatively obtain the spin excitation (multi-magnon) spectrum schematically depicted in Fig. 3. Two-magnon excitations with $K_x = k$ can be constructed from two single magnons of momenta $\frac{1}{2}k + k'$ and $\frac{1}{2}k - k'$, leading to a broad two-magnon continuum (except near $K_x = \pi$). There are two different types of two-magnon continua corresponding to the excitation of magnons of same or opposite parities w.r.t. $R_x$ (even-even, odd-odd, or even-odd excitations) as shown in Fig. 3.

We turn now to the magnetic susceptibility, $\chi(T)$, at low temperatures, $T \ll \Delta$, where magnon density is $\ll 1$ per rung and magnon-magnon interactions are negligible. The dominant term is $\propto e^{-\Delta/T}$, due to the finite gap, there is a prefactor determined by the dispersion of the one-magnon spectrum. At the isotropy ($J'=J$), it suffices to include only the lowest odd-parity ($B$) branch. If the dispersion has a form, $\Delta + bk^\nu$, near its minimum, the prefactor in $\chi$ is $\propto T^{1/\nu - 1}$. The finite size corrections of the spin gap in Ref.[8] indicate $\nu = 2$ and $b \sim 10/\pi^2 \sim 1$, leading to

$$\chi(T) \sim \frac{(g\mu_B)^2}{4\sqrt{\pi b T}} e^{-\Delta/T} \quad \text{(per site)}, \tag{4}$$

where $g$ is the $g$-factor and $\mu_B$ is the Bohr magneton.[13]

Holes doped in this spin liquid should, at least for some parameters, form bound pairs like in the single ladder[9] or 2D[14] cases; e.g. at small $J/J'$ and $t/J'$, two holes form a local pair on a bond to minimize the magnetic cost of $J'$. In this limit there are two kinds of triplet excitations[9]; (i) a local triplet excitation away from the hole pair and (ii) a breaking of the hole pair into a spin triplet combination of two "quasiparticles (QP)" with charge $+|e|$ and spin 1/2. The isotropic limit ($J=J'$) is more involved and one needs numerical calculations to determine the exact nature of the spin excitations. The lowest two-hole energies in the 4×6 cluster with $J=J'$ are shown in Fig. 4(a) calculated for $K_x=0$, and all possible point group symmetries. When $J/t$ is not too small, two singlet states are pulled out from the higher energy continuum. These states can be considered as B ($R_x=1$) and A ($R_x=-1$) two-hole bound pairs with opposite parity w.r.t. the $x$-axis, or alternatively assigned a momentum $K_y=0$ or $\pi$ in the transverse direction. The B–A splitting is proportional to the pair hopping amplitude along $y$, i.e. a fraction of $t$. In this regime the lowest triplet excitation has $R_x=1$ and $R_y=-1$ (see Fig. 4(a)). The corresponding spin gap is shown in Fig. 4(b). There is no sign of collapse of the spin gap which remains finite down to $J/t=0.17$. It should be noted that two-hole wave functions carrying finite total momenta along $x$ or $y$ (i.e. A combination) have increased energies of a few fractions of $t$. These are modes of coherent propagation of the bound hole pair. Hence, at sufficiently small $J/t$ ratio, some states at the top of the two-hole band (e.g. the A hole pair as shown in Fig. 4(b)) can lie above the bottom of the triplet excitation spectrum even though the spin gap is finite. In other words, the width of the hole-pair band is larger than the spin gap.

There are several triplet excitations with different symmetries: the lowest has parity $R_x=+1$, and there are others with $R_x=-1$. Based on their parity, these excitations are tentatively assigned as follows. The lowest is of type (ii), i.e., obtained by breaking the bound pair into two QP's with spin 1/2. The two QP's then occupy the "one-rung" GS at different rungs (more precisely, extended over a couple of rungs), and propagate along the chain direction mostly independently. The parity of the two-QP state is therefore the square of the parity of the one-rung orbital (actually $R_x=+1$), i. e. $R_x=+1$ as observed for the lowest triplet. Excitations of type (i), i.e. a triplet rung away from the bound hole pair, have a total parity given by the product of the parity of the two-hole singlet GS and that of the triplet rung. As seen in Figs. 2 and 4(a), the lowest triplet in the Heisenberg ladder has a parity $R_x=-1$ and the two-hole singlet GS has $R_x=+1$. Therefore, the lowest type (i) excitation excitation will have a parity $R_x=-1$. This mode will have an energy minimum at $K_x=\pi$, because the lowest excitation in the Heisenberg ladder is there. Therefore, the excitation energy of this mode is not directly related to the data of Fig. 2, since all the states in Fig. 4(a) have a total momentum $K_x=0$, corresponding to magnon with an arbitrary momentum $q$ plus a singlet hole-pair with momentum $-q$, while the rest of our data correspond only to magnons with $q=\pi$ and singlets with $q=0$. We note that the type (i) excitations should lie higher than the spin gap in the Heisenberg ladder due to charge fluctuations, which is actually the case in the single ladder.[9,12] The triplet mode with $R_x=-1$ and $R_y=-1$ has a similar excitation energy to the two-QP spin gap, and so probably is a two-QP excitation from another singlet state, i.e. with symmetries $R_x=-1$ and $R_y=+1$. This singlet state with $R_x=-1$ may be described by a "charge polarization" in the GS, and has an energy higher by $\approx 0.1t$. But the energy shift of the triplet excitation may be smaller, because of a residual attraction between the charge polarization and the two QP's. energy of this mode. Numerical calculation of correlation functions are needed for a definite identification.

In conclusion, performing exact diagonalization of small $4\times L$ clusters we have investigated the low-lying spin excitations of the $t$–$J$ double-ladder. At half-filling, finite size scaling gives an accurate determination of the spin gap. Both the triplet excitation band and the two-hole band are split by the inter-ladder coupling into bonding and anti-bonding combinations but the spin gap remains finite. Neglecting interactions between elementary low energy spin excitations we have also estimated qualitatively the full spin excitation spectrum and the temperature behavior of the spin susceptibility at half-filling.

The authors thank M. Troyer and R. Noack for useful discussions. *Laboratoire de Physique Quantique, Toulouse* is *Unité de Recherche Associé au CNRS No 505*. DP acknowl-edges support from the EEC Human Capital and Mobility program under grant CHRX-CT93-0332. We also thank IDRIS (Orsay) for allocation of CPU time on the CRAY-C98 computer. HT was supported by the Swiss National Science Foundation.

FIGURES

FIG. 1. Two coupled ladders. $J'$ corresponds to the magnetic coupling on the rungs of the two ladders (thick lines) and $J$ to the intra- and inter-ladder couplings (thin lines). Periodic boundary conditions are used in the chain direction. Symmetries $R_x$ and $R_y$ are shown on the plot.

FIG. 2. Triplet (or odd spin) excitation spectrum of the Heisenberg double ladder at momentum $\pi$ vs $J/J'$. Different symbols are associated to the 4 possible spatial symmetries as indicated on the plot (see text). The dashed line corresponds to a $2^{nd}$ order perturbation in $J/J'$ and the exact 4×4 data are shown by a full line.

FIG. 3. Schematic picture of the triplet excitation spectrum vs momentum. Thick solid and dashed lines correspond to the $B$ and $A$ one-magnon dispersions respectively. The hatched regions denote the two-magnon continuum. Excitations with more than two magnons are omitted. The extrema of the various continua corresponding to two magnons of same (opposite) parities are delimited by thin solid (dashed) lines.

FIG. 4. (a) Zero momentum two hole GS energies (in units of $t$) in various symmetry sectors in a 4×6 double ladder vs $J/t$. Correspondence between symbols and GS symmetries is shown on the plot. The energies are measured from the GS energy in the static limit ($t$=0). (b) Spin gap vs $J/t$ extracted from (a) corresponding to the lowest triplet excitation of the $B$ ($R_x$=1) two-hole bound pair (■). The energy difference between the same triplet excitation and the *excited A* ($R_x$=−1) two-hole pair is also shown (□). As a reference the spin gap of the Heisenberg 4×6 multi-ladder is indicated on the plot.

TABLES

TABLE I. Spin gaps $\Delta(L)$ in $4\times L$ double ladder clusters. The extrapolated value $\Delta(\infty)$ is obtained by a least square fit of the numerical data of the form $\Delta(L)\sim\Delta(\infty)+a\,\exp(-L/\xi)$. The corresponding fits for the parameters $a$ and $\xi$ are also given in the table.

| $L$ | 0.5 | 0.75 | 1.0 |
|---|---|---|---|
| 6 | 0.453716 | 0.364220 | 0.386821 |
| 8 | 0.440387 | 0.314560 | 0.308908 |
| $\infty$ | $0.436 \pm 0.004$ | $0.280 \pm 0.01$ | $0.245 \pm 0.02$ |
| a | $1.2 \pm 0.2$ | $1.3 \pm 0.1$ | $1.75 \pm 0.05$ |
| $\xi$ | $1.45 \pm 0.15$ | $2.2 \pm 0.1$ | $2.4 \pm 0.1$ |